\documentclass[aps,pre,reprint,onecolumn,notitlepage]{revtex4-1}
\usepackage[english]{babel}
\usepackage{amsmath, amsthm, amssymb, latexsym, graphicx}
\usepackage[utf8]{inputenc}
\usepackage[usenames,dvipsnames]{xcolor}

\newcommand{\la}{\lambda}
\newcommand{\ome}{\boldsymbol{\omega}_{e}}
\newcommand{\omp}{\boldsymbol{\omega}_{p}}
\newcommand{\ue}{\textbf{u}_{e}}
\newcommand{\up}{\textbf{u}_{p}}

\begin{document}

\title{Exact scaling laws for helical three-dimensional two-fluid turbulent plasmas}

\author{N. Andr\'es$^{1}$}
\author{S. Galtier$^{1,2}$}
\author{F. Sahraoui$^{1}$}
\affiliation{$^1$ LPP, \'Ecole Polytechnique, F-91128 Palaiseau Cedex, France \\
             $^2$ Departement de Physique, Universit\'e Paris-Sud, Orsay, France.}
\date{\today}

\begin{abstract}
We derive exact scaling laws for a three-dimensional incompressible helical two-fluid plasma, without the assumption of isotropy. For each ideal invariant of the two-fluid model, i.e. the total energy, the electron helicity and the proton helicity, we derive simple scaling laws in terms of two-point increment correlation functions expressed in terms of the velocity field of each species and the magnetic field. These variables are appropriate for comparison {with direct numerical simulation data and with \textit{in-situ} measurements in the near Earth space over a broad range of spatial scales} . Finally, using the exact scaling laws and dimensional analysis we predict the magnetic energy and electron helicity spectra for different ranges of scales.
\end{abstract}

\maketitle

\section{Introduction}\label{intro}

The first exact law for incompressible hydrodynamic turbulence is the so-called von K\'arm\'an-Howarth equation, which relates the time evolution of the second-order correlation velocity tensor to the divergence of the third-order correlation velocity tensor \citep{vKH1938}. Under the assumption of isotropy and homogeneity, \citet{vKH1938} found this exact result, which is considered as one of the cornerstones of turbulence theories \citep[e.g.][]{F1995}. The most important consequence of the von K\'arm\'an-Howarth equation is the “four-fifth” law, which predicts a linear scaling for the longitudinal two-point third-order velocity structure function with the distance between the two points. As a consequence, this exact scaling law puts strong constraints on the nonlinear dynamics of turbulent flows. 

Following a similar formalism of \citet{vKH1938}, other exact laws have been derived under the assumptions of homogeneity and isotropy of the turbulent fluctuations for helical hydrodynamic \citep{Ch1996,K2003}, magnetohydrodynamic (MHD) \citep{Ch1951,P1998a,P1998b}, helical MHD \citep{P2003}, Hall-MHD (HMHD) \citep{G2008}, electron-MHD approximation (EMHD) \citep{Me2010} and Lagrangian averaged models \citep{H2002,P2006}. Recently, \citet{A2016b} have derived the von K\'arm\'an-Howarth equation for a 3D incompressible two-fluid plasma and the equivalent of the hydrodynamic four-fifth law. However, in all those derivations the assumption of isotropy limits the applicability of the laws, in particular, in magnetized (space or laboratory) plasmas. Indeed, in those plasmas, e.g. the solar wind, the presence of a non-zero mean magnetic field influences the statistical properties of the turbulence such as the spatial correlation function~\citep[e.g.][]{M1995}. This results in having different scaling properties along and perpendicular to the local mean field~\citep{Sh1983,Mu2009}. Therefore, general exact laws that go beyond the assumption of spatial isotropy are needed to study the nonlinear dynamics in turbulent space plasmas, in particular, when comparing to \textit{in-situ} spacecraft obervations.

\citet{Ga2011} have derived an exact law for the two-point correlation function associated with the total energy in 3D compressible isothermal hydrodynamic (HD) turbulence, without the assumption of isotropy. The authors found the presence of new types of terms in the inertial range (other than the Yaglom-like flux terms), which play the role of sources or sinks for the mean energy transfer rate. In the same line of research, they derived an exact law for compressible isothermal MHD turbulence using the two-point correlation function associated with the total energy \citep{B2013}. Recent works have studied scaling laws for correlation functions associated with the total energy \citep{B2016b} and the magnetic (and generalized) helicity \citep{B2016} for the HMHD model. It is worth mentioning that these exact scaling laws give accurate estimates of the mean transfer rate (of the particular ideal invariant), which is an essential quantity to characterize a turbulent system. In the present paper, we derive exact scaling laws for the two-point correlation functions associated with each ideal invariant of a 3D incompressible and homogeneous two-fluid plasma, without the assumption of isotropy.

The two-fluid model used in this work is derived from the general two-fluid equations when the non-relativistic and the quasi-neutrality approximation are used. These two assumptions correspond to neglecting the displacement current in the Maxwell-Amp\`ere equation~\citep{S2003}, which filters out the three high frequency (optic) eigenmodes of the general two-fluid equations. The resulting (reduced) two-fluid model still retains small scale effects that are the Hall term and electron inertia. The incompressibility assumption is further used; thus the resulting model does not include either compressible modes or finite Larmor radius effects. In the linear limit, the system supports two propagating modes at high frequency, generally referred to as whistler (right-handed) and Alfv\'en mode (left-handed), which become degenerate in the MHD limit. The retained (Hall and electron inertia) terms introduce new spatial and temporal scales into the theoretical plasma description \citep[e.g.][]{Bi1997}, which are respectively the ion and electron gyrofrequencies and inertial lengths. It is worth mentioning that the two-fluid description includes MHD, HMHD and EMHD models, which can be regarded as particular cases in the proper asymptotic limits. For instance, at length scales larger than the ion inertial length, the Hall effect and electron inertia can be neglected. At those largest scales, the MHD description is appropriate. At spatial scales comparable or smaller than the ion-skin depth, the ions are no longer frozen-in to the magnetic field lines because of the Hall term. At those intermediate scales, the HMHD description becomes valid and has been extensively studied both numerically \citep{Mi2003,Ma2003,D2006,Ga2007,M2012} and analytically \citep{T1986,MY1998,Mi2002,S2003,S2007,B2016}. In the very high frequency limit of the two-fluid model, ions can be considered as motionless (because of their large mass with respect to electrons) and provide a neutralizing background, while the electrons carry the full electric current. This approximation corresponds to the EMHD model, and is asymptotically valid at spatial scales comparable or smaller than electron inertial length. In conclusion, the two-fluid description used in this work retains the whole dynamics of both the ion and electron flows from the MHD down to the electron inertial length scales, within the non-relativistic, quasi-neutrality and incompressibility approximations.

Using a recent alternative formulation \citep{B2001b}, we derive exact scaling laws for the three ideal invariants of a 3D incompressible and homogeneous two-fluid plasma. The rest of the paper is organized as follow: {in Section \ref{model} we introduce the 3D incompressible two-fluid model and in} Section \ref{ideal_inv} its ideal invariants. In Section \ref{results} we present our main theoretical results for each ideal invariant, namely the total energy, the electron helicity and the proton helicity. In Section \ref{disc} we discuss the implications of the derived exact scaling laws, and finally, {in Section \ref{conclus}} we provide a summary of the results.
%-------------------------------------------------------------
\section{Two fluid equations}\label{model}
%-------------------------------------------------------------
The equations of motion for a quasi-neutral incompressible plasma of ions and electrons with masses $m_{i,e}$, charges $\pm e$, constant densities $n_{p}=n_{e}=n$, pressures $p_{e,p}$, and respective velocities $\ue$ and $\up$ are \citep{A2016}
\begin{align}\label{eq:1}
m_en\frac{d \ue}{dt} &=  -en\left(\textbf{E}+\frac{1}{c}\ue\times\textbf{B}\right)-\boldsymbol\nabla p_{e} + \textbf{f}_e + \textbf{d}_e, \\ \label{eq:2} 
m_in\frac{d \up}{dt} &=  en\left(\textbf{E}+\frac{1}{c}\up\times\textbf{B}\right)-\boldsymbol\nabla p_{p} + \textbf{f}_p + \textbf{d}_p, \\ 
\textbf{J} &= \frac{c}{4\pi}\boldsymbol\nabla\times\textbf{B}={en}(\up-\ue) .
\end{align}
Here $d/dt = \partial/\partial t + {\bf u} \cdot \nabla$ is the total derivative, $\textbf{B}$ and $\textbf{E}$ are the magnetic and electric fields, $\textbf{J}$ is the electric current density, $c$ is the speed of light and $\textbf{f}_{e,p}$ and $\textbf{d}_{e,p}$ are the forcing and dissipative terms, respectively. Note that these fluid equations do not include any kinetic plasma dissipation mechanisms (e.g., wave-particle interactions) due to either electrons or ions.

The incompressibility assumption implies,
\begin{align}
\boldsymbol\nabla\cdot\textbf{u}_e &= 0, \\
\boldsymbol\nabla\cdot\textbf{u}_p &= 0.
\end{align}
Equations (\ref{eq:1}) and (\ref{eq:2}) can be written in dimensionless form in terms of a typical length $L_0$, the particle density $n$, a typical velocity $v_A=B_0/(4\pi nM)^{1/2}$ (the Alfv\'en velocity, where $B_0$ is a typical value of ${\bf B}$, and $M\equiv m_i+m_e$), and with the electric field in units of $E_0 = v_AB_0/c$,
\begin{align}\label{dlesse}
 \mu\frac{d \ue}{dt} &= -\frac{1}{\lambda}(\textbf{E}+\ue\times\textbf{B})-\boldsymbol\nabla p_{e} +  \textbf{f}_e + \textbf{d}_e, \\ 
\label{dlessp}
 (1-\mu)\frac{d \up}{dt} &= \frac{1}{\lambda}(\textbf{E}+\up\times\textbf{B})-\boldsymbol\nabla p_{p} + \textbf{f}_e + \textbf{d}_e, \\
\label{dlesso}
\textbf{J} &= \frac{1}{\lambda}(\up-\ue),
\end{align} 
where we have introduced the dimensionless parameters $\mu\equiv m_e/M$ and $\lambda\equiv c/(\omega_{M}L_0)$, where $\omega_{M}=(4\pi e^2n/M)^{1/2}$  has the form of a plasma frequency for a particle of mass $M$. Dimensionless ion and electron skin-depth can be defined in terms of their corresponding plasma frequencies $\omega_{i,e}=(4\pi e^2n/m_{i,e})^{1/2}$ simply as $\lambda_{i,e}\equiv c/(\omega_{i,e}L_0)$, and their expressions in terms of $\mu$ and $\lambda$ {are} $\lambda_i=(1-\mu)^{1/2}\lambda$ and $\lambda_e=\mu^{1/2}\lambda$. Note that in the limit of electron inertia equal to zero, we obtain $\lambda = \lambda_i = c/(\omega_{i}L_0)$, which correspond to the usual Hall parameter \citep{T1986}. Finally, to obtain a hydrodynamic description of the two-fluid plasma, we can write \textbf{u}$_e$ and \textbf{u}$_p$ in terms of two vector fields (see \cite{A2014a}): the hydrodynamic velocity $\textbf{U} = (1-\mu)\textbf{u}_p+\mu\textbf{u}_e$, and $\textbf{J}$ as given by Eq.~\eqref{dlesso}. From these two fields, it is trivial to obtain \textbf{u}$_e$ and \textbf{u}$_p$ as 
\begin{align}\label{ue}
\textbf{u}_e&=\textbf{U} - (1-\mu)\lambda\textbf{J}, \\\label{up}
\textbf{u}_p&=\textbf{U} + \mu\lambda\textbf{J}.
\end{align}
%---------------------------------------------------------------------------
 \section{Ideal invariants}\label{ideal_inv}
%---------------------------------------------------------------------------
In general, a multi-fluid plasma made of $N$ species has $N+1$ ideal invariants. For a 3D incompressible two-fluid plasma, using $\textbf{E}=-\partial_t\textbf{A}-\boldsymbol\nabla\phi$, we can readily show that the total energy $E_T$ is one of these ideal invariants, where
\begin{equation}\label{ene}
 E_T = \frac{1}{2}\int d^3r \bigg(\mu u^{2}_e+ (1-\mu)u^{2}_p+B^2\bigg).
\end{equation}
The other two invariants are the electron helicity and the proton helicity,
\begin{align}\label{hele}
H_{e} &= \frac{1}{2}\int d^3r \bigg(\textbf{A}-\lambda\mu\ue\bigg)\cdot\bigg(\textbf{B}-\lambda\mu\ome\bigg), \\ \label{help}
H_{p} &= \frac{1}{2}\int d^3r \bigg(\textbf{A}+\lambda(1-\mu)\up\bigg)\cdot\bigg(\textbf{B}+\lambda(1-\mu)\omp\bigg), 
\end{align}
where $\boldsymbol\omega_{e,p}=\boldsymbol\nabla\times\textbf{u}_{e,p}$. If we define the electron and proton vector potentials
\begin{align}\label{pote}
\textbf{h}_{e} &= \textbf{A}-\la\mu\ue, \\
\label{potp}
\textbf{h}_{p} &= \textbf{A}+\la(1-\mu)\up,
\end{align}
equations \eqref{hele} and \eqref{help} can be casted in a compact expressions as 
\begin{align}\label{helie}
{H}_{e} &= \int dr^3~\textbf{h}_{e}\cdot\textbf{H}_{e}, \\ \label{helip}
{H}_{p} &= \int dr^3~\textbf{h}_{p}\cdot\textbf{H}_{p},
\end{align}
where
\begin{align}
\textbf{H}_{e} &= \boldsymbol\nabla\times\textbf{h}_{e} = \textbf{B}-\la\mu\ome, \\
\textbf{H}_{p} &= \boldsymbol\nabla\times\textbf{h}_{p} = \textbf{B}+\la(1-\mu)\omp.
\end{align}
It is worth mentioning that in the HMHD limit, i.e. $\mu\rightarrow0$ and $\lambda\rightarrow\lambda_i$, the conservation of the electron helicity and proton helicity corresponds to the conservation of the magnetic helicity and generalized helicity, respectively \citep{W1958,T1986}.
%-------------------------------------------------------------
\section{Exact scaling laws}\label{results}
%-------------------------------------------------------------

\subsection{The total energy $E_T$}\label{total_energy}

Following recent works \citep{Ga2011,B2016}, we define the symmetric two-point correlation functions associated with the energy of each species as,
\begin{align}
R_{E_e} &= R_{E_e}' = \frac{1}{2} \left<\ue\cdot\ue'\right>, \\
R_{E_p} &= R_{E_p}' = \frac{1}{2}\left<\up\cdot\up'\right>,
\end{align}
where the prime denotes field evaluation at $\textbf{x}'=\textbf{x}+\textbf{r}$ (being \textbf{r} the displacement vector) and the angular bracket denotes an ensemble average. The property of spatial homogeneity implies that all regions of space are similar so far as the statistical properties are concerned, which suggests that the results of averaging over a large number of realizations at different positions in space could be obtained equally well by averaging over a large region of space for one realization \citep{Ba1953}. 

Using equations \eqref{dlesse} and \eqref{dlessp}, with the corresponding large-scale forcing terms $\textbf{f}_{e,p}$ and small-scales dissipation terms $\textbf{d}_{e,p}$ in each equation (since we expect a direct cascade for the total energy \citep{M1982}), we obtain a time evolution for the symmetric two-point correlation function associated with the total kinetic energy $R_{E_k}\equiv R_{E_e}+R_{E_p}$ as,
\begin{align}\nonumber
\frac{\partial}{\partial t}\left<R_{E_k}+R_{E_k}'\right> &=~ \mu\left<\ue'\cdot\frac{\partial\ue}{\partial t}+\ue\cdot\frac{\partial\ue'}{\partial t}\right>+(1-\mu)\left<\up'\cdot\frac{\partial\up}{\partial t}+\up\cdot\frac{\partial\up'}{\partial t}\right> \\\nonumber
&=\left<\ue'\cdot\big[\ue\times(\mu\ome-\frac{1}{\lambda}\textbf{B})\big]\right>+\left<\ue\cdot\big[\ue'\times(\mu\boldsymbol{\ome}'-\frac{1}{\lambda}\textbf{B}')\big]\right> \\\nonumber 
&+\left<\up'\cdot\big\{\up\times[(1-\mu)\omp+\frac{1}{\lambda}\textbf{B}]\big\}\right>+\left<\up\cdot\big\{\up'\times[(1-\mu)\omp'+\frac{1}{\lambda}\textbf{B}']\big\}\right> \\\nonumber
&+\left<\textbf{J}'\cdot\textbf{E}+\textbf{J}\cdot\textbf{E}'\right> - \left<\ue'\cdot\boldsymbol\nabla P_{e} - \ue\cdot\boldsymbol\nabla' P_{e}' - \up'\cdot\boldsymbol\nabla P_{p} - \up\cdot\boldsymbol\nabla' P_{p}'\right> \\ \label{totene}
&+ \mathcal{D} + \mathcal{F}
\end{align}
where we have used equations \eqref{ue} and \eqref{up}, {i.e. the expressions for the velocity field of electrons and protons}, and we have defined $P_{e,p}\equiv \lambda p_{e,p}+s_{e,p}\lambda u_{e,p}^2/2$, with $s_e=\mu$ and $s_p=1-\mu$, {and $\mathcal{D}$ and $\mathcal{F}$ are given by} 
\begin{align}\nonumber
\mathcal{D} &=~ \left<\textbf{d}_e\cdot\ue'+\textbf{d}_e'\cdot\ue+\textbf{d}_p\cdot\up'+\textbf{d}_p'\cdot\up\right>,\\
\mathcal{F} &=~ \left<\textbf{f}_e\cdot\ue'+\textbf{f}_e'\cdot\ue+\textbf{f}_p\cdot\up'+\textbf{f}_p'\cdot\up\right>.
\end{align}
Using the vectorial property $\boldsymbol\nabla\cdot(\textbf{a}\times\textbf{b})=(\boldsymbol\nabla\times\textbf{a})\cdot\textbf{b}-(\boldsymbol\nabla\times\textbf{b})\cdot\textbf{a}$ and the homogeneity {assumption}, we can readily obtain an expression for the symmetric two-point correlation function associated with the magnetic energy
\begin{align}
R_{E_B} = \frac{1}{2} \left<\textbf{B}\cdot\textbf{B}'\right>,
\end{align}
since 
\begin{align}\nonumber
\left<\textbf{J}'\cdot\textbf{E}+\textbf{J}\cdot\textbf{E}'\right>&=~\left<(\boldsymbol\nabla\times\textbf{B}')\cdot\textbf{E}+\boldsymbol\nabla\times\textbf{B}\cdot\textbf{E}'\right> \\\nonumber
&=~\left<\boldsymbol\nabla'\cdot(\textbf{B}'\times\textbf{E})+\boldsymbol\nabla\cdot(\textbf{B}\times\textbf{E}')\right> \\\nonumber
&=~-\left<\boldsymbol\nabla\cdot(\textbf{B}'\times\textbf{E})-\boldsymbol\nabla'\cdot(\textbf{B}\times\textbf{E}')\right> \\\nonumber
&=~\left<(\boldsymbol\nabla\times\textbf{E})\cdot\textbf{B}'+(\boldsymbol\nabla'\times\textbf{E}')\cdot\textbf{B}\right> \\\nonumber
&=-\left<\textbf{B}'\cdot\frac{\partial\textbf{B}}{\partial t}+\textbf{B}\cdot\frac{\partial\textbf{B}'}{\partial t}\right> \\\nonumber
&= -\frac{\partial}{\partial t}\left<R_{E_{B}}+R_{E_{B}}'\right>.
\end{align}
Then, using definitions \eqref{pote} and \eqref{potp} and defining the symmetric two-point correlation function associated with the total energy as $R_{E_T}\equiv R_{E_k}+R_{E_B}$, equation \eqref{totene} can be written as,
\begin{align}\nonumber
\frac{\partial}{\partial t}\left<R_{E_T}+R_{E_T}'\right> &=-\frac{1}{\lambda}\left<\ue'\cdot(\ue\times\textbf
{H}_e)+\ue\cdot(\ue'\times\textbf{H}_e')\right> \\ \label{totene2}
&+\frac{1}{\lambda}\left<\up'\cdot(\up\times\textbf{H}_p) + \up\cdot(\up'\times\textbf{H}_p')\right> + \mathcal{D} + \mathcal{F}
\end{align}
where the terms involving a gradient vanish by incompressibility and homogeneity. Equation \eqref{totene2} is an exact law for a incompressible two-fluid plasma, even in anisotropic turbulence \citep{M1981,C2009}. Assuming the existence of an inertial energy range, in the limit of infinite Reynolds numbers ($\mathcal{D}\to0$) and considering a statistical stationary regime ($\partial_t\sim0$), we obtain
\begin{align}\label{totene3}
2\varepsilon_{E_T} &=\frac{1}{\lambda}\big[\left<\ue'\cdot(\ue\times\textbf
{H}_e) + \ue\cdot(\ue'\times\textbf{H}_e')\right> -\left<\up'\cdot(\up\times\textbf{H}_p) + \up\cdot(\up'\times\textbf{H}_p')\right>\big],
\end{align}
where $\mathcal{F}=2\varepsilon_{E_T}$, with $\varepsilon_{E_T}$ the mean energy dissipation rate per unit mass. Finally, for a given field \textbf{a}, we introduce the two-point increment correlation function as $\delta\textbf{a}\equiv\textbf{a}'-\textbf{a}$. Therefore, equation \eqref{totene3} can be written as,
\begin{align}\label{totene4}
2\varepsilon_{E_T} &=\frac{1}{\lambda}[\left<\delta(\up\times\textbf{H}_p)\cdot\delta\up\right>-\left<\delta(\ue\times\textbf
{H}_e)\cdot\delta\ue\right>]
\end{align}
{where we have used the property that $\textbf{u}_{e,p}$ is perpendicular to $\textbf{u}_{e,p}\times\textbf{H}_{e,p}$.} The exact scaling law \eqref{totene4} is our first main result. It is worth mentioning that this law is valid only in the inertial range and, therefore, is independent of the dissipation mechanisms present in the plasma (assuming that the dissipation terms act only at the largest wavenumbers). Furthermore, equation \eqref{totene4} is written only in terms of the two-point increment correlation functions, which are {rather easy} to obtain from \textit{in-situ} measurements and data from numerical simulations. In particular, these two-point increments are written as a function of the velocity of each species and the magnetic field, variables which {now can be measured down to the electron scales in the near-Earth space by the recently launched NASA/MMS (Magnetospheric Multiscale) mission \citep{Bu2016}}. {Equation \eqref{totene4} is made of two terms, one per each species. The main contribution to this scaling law is the proton term, which is mainly responsible for the dynamics at the large scales and to a lesser extent at intermediate and small scales. On the other hand, the electron term mainly contributes to the smallest scales through the terms proportional to $\mu$. Finally, it is worth mentioning that} in the HMHD ($\mu\rightarrow0$, $\lambda\rightarrow\lambda_p$) and MHD limits ($\mu\rightarrow0$ and $\lambda\rightarrow0$) we recover the result reported {in} \citet{B2016b}.
%-----------------------------------------------------------------------------------------
\subsection{Electron helicity $H_e$ and proton helicity $H_p$}\label{helicities}
%-----------------------------------------------------------------------------------------
The equation of motion for electrons and protons \eqref{dlesse} and \eqref{dlessp}, using $\textbf{E}=-\partial_t\textbf{A}-\boldsymbol\nabla\phi$, and assuming the existence of small-scale forcing $\bar{\textbf{f}}_{e,p}$ (since we expect an inverse cascade for each helicity \citep{P1957,G2015}), can be casted into 
\begin{align}\label{dlesse2}
 \frac{\partial}{\partial t}(\textbf{A}-\la\mu\ue) &= \ue\times(\textbf{B}-\la\mu\ome) - \boldsymbol\nabla \bar{P}_{e} + \bar{\textbf{d}}_e + \bar{\textbf{f}}_e , \\
\label{dlessp2}
 \frac{\partial}{\partial t}[\textbf{A}+\la(1-\mu)\up] &= \up\times[\textbf{B}+\la(1-\mu)\omp]-\boldsymbol\nabla \bar{P}_{p} + \bar{\textbf{d}}_p +\bar{\textbf{f}}_p
\end{align} 
where we have defined $\bar{P}_{e,p} = P_{e,p}+\phi$ and $\bar{\textbf{d}}_{e,p}$ is the large-scale dissipation. This last term is introduced to prevent the formation of a condensate state \citep{C2007}. Similar models have been studied in the literature using this technique \citep{K1967}. Using definitions \eqref{pote} and \eqref{potp}, equations \eqref{dlesse2} and \eqref{dlessp2} can be written as
\begin{align}\label{dlesse3}
 \frac{\partial\textbf{h}_{e}}{\partial t} &= \ue\times\textbf{H}_{e} - \boldsymbol\nabla P_{e} + \bar{\textbf{f}}_e + \bar{\textbf{d}_e}, \\
\label{dlessp3}
 \frac{\partial\textbf{h}_{p}}{\partial t} &= \up\times\textbf{H}_{p} - \boldsymbol\nabla P_{p} + \bar{\textbf{f}}_p + \bar{\textbf{d}_p}.
\end{align}
For the computation of the exact scaling for each helicity, we use the curl of equations \eqref{dlesse3} and \eqref{dlessp3},
\begin{align}\label{dlesse3p}
\frac{\partial \textbf{H}_e }{ \partial t} &= \boldsymbol\nabla\times(\textbf{u}_e\times\textbf{H}_e) + \bar{\textbf{F}}_e + \bar{\textbf{D}_e}, \\ \label{dless3p}
\frac{\partial \textbf{H}_p }{ \partial t} &= \boldsymbol\nabla\times(\textbf{u}_p\times\textbf{H}_p)+ \bar{\textbf{F}}_p + \bar{\textbf{D}}_p
\end{align}
where $\bar{\textbf{F}}_{e,p}=\boldsymbol\nabla\times\bar{\textbf{f}}_{e,p}$ and $\bar{\textbf{D}}_{e,p}=\boldsymbol\nabla\times\bar{\textbf{d}}_{e,p}$. As {in Section \ref{total_energy}}, we define the symmetric two-point correlation function associated with the helicity of each species as,
\begin{align}
R_{E_{He}} &= R_{E_{He}}' = \frac{1}{2} \left<\textbf{H}_e\cdot\textbf{h}_e'+\textbf{h}_e\cdot\textbf{H}_e'\right>, \\
R_{E_{Hp}} &= R_{E_{He}}' = \frac{1}{2} \left<\textbf{H}_p\cdot\textbf{h}_p'+\textbf{h}_p\cdot\textbf{H}_p'\right>.
\end{align}
Using equations \eqref{dlesse3}-\eqref{dless3p}, {which are the equations of motion (and its curl) for electrons and protons written in terms of $\textbf{h}_{e,p}$ and $\textbf{H}_{e,p}$,} we can obtain a time evolution for the symmetric two-point correlation functions as,
\begin{align}\nonumber
\frac{\partial}{\partial t}\left<R_{E_{H_e}}+R_{E_{H_e}}'\right> &=~ \left<\textbf{H}_{e}'\cdot\frac{\partial\textbf{h}_{e}}{\partial t}+ \textbf{H}_{e}\cdot\frac{\partial\textbf{h}_{e}'}{\partial t}+\textbf{h}_{e}'\cdot\frac{\partial\textbf{H}_{e}}{\partial t}+\textbf{h}_{e}\cdot\frac{\partial\textbf{H}_{e}'}{\partial t}\right> \\ \nonumber &=~\left<\textbf{H}_{e}'\cdot(\ue\times\textbf{H}_{e})\right> + \left<\textbf{H}_{e}\cdot(\ue'\times\textbf{H}_{e}')\right> \\
&~+\left<\textbf{h}_{e}\cdot\boldsymbol\nabla'\times(\ue'\times\textbf{H}_{e}')\right>+\left<\textbf{h}_{e}'\cdot\boldsymbol\nabla\times(\ue\times\textbf{H}_{e})\right> +~ \bar{\mathcal{D}}_e + \bar{\mathcal{F}}_e, \\ \nonumber
\frac{\partial}{\partial t}\left<R_{E_{H_p}}+R_{E_{H_p}}'\right> &=~ \left<\textbf{H}_{p}'\cdot\frac{\partial\textbf{h}_{p}}{\partial t}+\textbf{H}_{p}\cdot\frac{\partial\textbf{h}_{p}'}{\partial t}+\textbf{h}_{p}'\cdot\frac{\partial\textbf{H}_{p}}{\partial t}+\textbf{h}_{p}\cdot\frac{\partial\textbf{H}_{p}'}{\partial t}\right>~ \\ \nonumber &=~\left<\textbf{H}_{p}'\cdot(\up\times\textbf{H}_{p})\right> + \left<\textbf{H}_{p}\cdot(\up'\times\textbf{H}_{p}')\right> \\
&~+\left<\textbf{h}_{p}\cdot\boldsymbol\nabla'\times(\up'\times\textbf{H}_{p}')\right>+\left<\textbf{h}_{p}'\cdot\boldsymbol\nabla\times(\up\times\textbf{H}_{p})\right> +~ \bar{\mathcal{D}}_p + \bar{\mathcal{F}}_p,
\end{align}
where again the gradient terms vanish by the incompressibility and homogeneity of the plasma. Using these conditions, we can show that,
\begin{align}\nonumber
\left<\textbf{h}_{e,p}\cdot\boldsymbol\nabla'\times(\textbf{u}_{e,p}'\times\textbf{H}_{e,p}')\right> &=~\boldsymbol\nabla'\cdot[(\textbf{u}_{e,p}'\times\textbf{H}_{e,p}')\times\textbf{h}_{e,p}] \\ &= -\boldsymbol\nabla\cdot[(\textbf{u}_{e,p}'\times\textbf{H}_{e,p}')\times\textbf{h}_{e,p}] = (\textbf{u}_{e,p}'\times\textbf{H}_{e,p}')\cdot\textbf{H}_{e,p}, \\ \nonumber
\left<\textbf{h}_{e,p}'\cdot\boldsymbol\nabla\times(\textbf{u}_{e,p}\times\textbf{H}_{e,p})\right> &=~\boldsymbol\nabla\cdot[(\textbf{u}_{e,p}\times\textbf{H}_{e,p})\times\textbf{h}_{e,p}'] \\ &= -\boldsymbol\nabla'\cdot[(\textbf{u}_{e,p}\times\textbf{H}_{e,p})\times\textbf{h}_{e,p}'] = (\textbf{u}_{e,p}\times\textbf{H}_{e,p})\cdot\textbf{H}_{e,p}'.
\end{align}
Therefore, introducing the two-point increments, we obtain the dynamical equations for each helicity as,
\begin{align}\nonumber
\frac{1}{2} \frac{\partial}{\partial t}\left<R_{E_{H_e}}+R_{E_{H_e}}'\right> &=~\left<\textbf{H}_e'\cdot(\textbf{u}_e\times\textbf{H}_e)\right>+\left<\textbf{H}_e\cdot(\textbf{u}_e'\times\textbf{H}_e')\right> +~ \frac{\bar{\mathcal{D}}_e}{2} + \frac{\bar{\mathcal{F}}_e}{2} \\ \label{helie2}
&= -\left<\delta(\textbf{u}_e\times\textbf{H}_e)\cdot\delta\textbf{H}_e\right> +~ \frac{\bar{\mathcal{D}}_e}{2} + \frac{\bar{\mathcal{F}}_e}{2}, \\ \nonumber
\frac{1}{2} \frac{\partial}{\partial t}\left<R_{E_{H_p}}+R_{E_{H_p}}'\right> &=~\left<\textbf{H}_p'\cdot(\textbf{u}_p\times\textbf{H}_p)\right>+\left<\textbf{H}_p\cdot(\textbf{u}_p'\times\textbf{H}_p')\right> +~ \frac{\bar{\mathcal{D}}_p}{2} + \frac{\bar{\mathcal{F}}_p}{2} \\ \label{helip2}
&= -\left<\delta(\textbf{u}_p\times\textbf{H}_p)\cdot\delta\textbf{H}_p\right> +~ \frac{\bar{\mathcal{D}}_p}{2} + \frac{\bar{\mathcal{F}}_p}{2}.
\end{align}
Equations \eqref{helie2} and \eqref{helip2} are exact expressions for helical incompressible two-fluid plasmas. Assuming the existence of an inertial range far away from the forcing scales ($\bar{\mathcal{F}}_{e,p}\sim 0$), under quasi-stationary statistical conditions ($\partial_t\sim 0$) we obtain,
\begin{align}\label{helie3}
2\varepsilon_{H_e} &=~ \left<\delta(\textbf{u}_e\times\textbf{H}_e)\cdot\delta\textbf{H}_e\right>, \\ \label{helip3}
2\varepsilon_{H_p} &=~ \left<\delta(\textbf{u}_p\times\textbf{H}_p)\cdot\delta\textbf{H}_p\right>, 
\end{align}
{where we have used $\bar{\mathcal{D}}_{e,p} = 4\varepsilon_{H_e,H_p}$ and $\varepsilon_{H_e,H_p}$ are the electron and proton helicity dissipation rates per unit mass}. Equations \eqref{helie3} and \eqref{helip3} are the second main result of the paper. These expressions are valid in the inertial range, without the assumption of isotropy. In particular, these results could be useful in astrophysical contexts where the condition of isotropy is usually not fulfilled \citep{H2010,GS1995}. Finally, as equation \eqref{totene4}, expressions \eqref{helie3} and \eqref{helip3} are written {as scalar products of fields increment correlation functions.} 
%---------------------------------------------------------
\section{Discussion}\label{disc}
%---------------------------------------------------------
Equations \eqref{totene4}, \eqref{helie3} and \eqref{helip3} are the main results of the present paper. These equations give exact relations for the two-point increments of anisotropic turbulence in an incompressible and homogeneous two-fluid plasma. In particular, these expressions give exact scaling relations for the three ideal invariants, i.e. the total energy, the electron helicity and the proton helicity. In contrast to previous results in the literature \citep{P1998b,P2003,Ga2008,B2016}, our results are written as a function of the velocity field of each species of the plasma and of the magnetic field. These quantities are directly measurable \textit{in-situ} in the near-Earth space, which should make straigthforward the estimation of the transfer rate of each invariant from spacecraft data \citep[e.g.][]{C2009,Ma2012} as well as from numerical simulation \citep[e.g.][]{Mi2009,W2010}. Moreover, since we retain the electron inertia, we are able to study the turbulence cascade from the MHD scales {down to the electron inertial scale length}. It is worth recalling that this broad range of scales cannot be captured by the HMHD or the massless EMHD models. Finally, our exact scaling laws are independent of the dissipation mechanism present in the plasma, since it only requires that the dissipation term gets off all the power injected by the forcing term at the very large scale. 

In the limit of large and intermediate scales, i.e. the MHD and HMHD ranges, we recover the exact laws recently reported by \citet{B2016b} for the total energy. Assuming isotropy and equipartition between magnetic and kinetic energy, expression \eqref{totene4} can be used to provide theoretical predictions for the magnetic energy spectrum in a turbulent plasma. {In fact, in a stationary and isotropic turbulent regime, the energy cascade corresponds to a constant energy flux in Fourier space $F_k$ which is therefore equal to the energy dissipation rate $\varepsilon$. For instance, in the case of incompressible hydrodynamic turbulence, the modulus of the energy flux in Fourier space goes like $F_k \sim k u^3_k = \varepsilon$, which leads to the well known Kolmogorov’s energy power spectrum $E_k \sim \varepsilon^{2/3} k^{-5/3}$, using $E_k \sim u^2_k / \tau_k$ and $\tau_k \sim (ku_k)^{-1}$ ($ \tau_k$ is the  nonlinear transfer time). At MHD scales ($k\ll\lambda_i^{-1}$) we recover the Kolmogorov spectrum, $E_B(k)\sim B_k^2/k\sim k^{-5/3}$ \citep{M1982,SV2007}, using the transfer nonlinear time $\tau_k\sim(kB_k)^{-1}$.} At HMHD scales ($k\sim\lambda_i^{-1}$ and $k\ll\lambda_e^{-1}$), using $\tau_k\sim(\lambda_ik^2B_k)^{-1}$, we obtain a magnetic spectrum $E_B(k)\sim k^{-7/3}$, which is roughly compatible with solar wind observations \citep{A2009,A2012,S2009,S2010,S2011,S2013} and numerical simulation results \citep{Mi2007,G2008,A2016b} at these intermediate scales. Finally, at the smallest scales ($k\sim\lambda_e^{-1}$), where we can assume than the proton motion is negligible with respect to the electron motion, the transfer nonlinear time $\tau_k\sim(\mu\lambda^3 k^4B_k)^{-1}$ leads to $E_B(k)\sim k^{-11/3}$. This scaling {has} been observed recently in numerical simulations \citep{A2014b} and is compatible with previous theoretical calculation in the EMHD approximation \citep{Me2010,Ab2016}. 

\begin{figure}
\centering
\includegraphics[width=.45\textwidth]{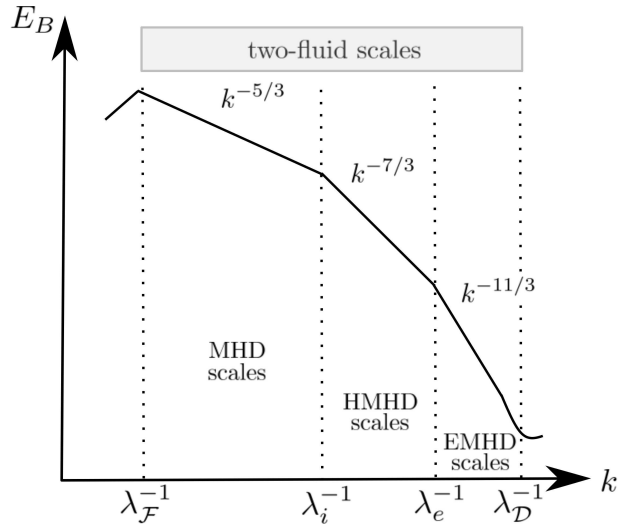} 
\caption{Schematic magnetic energy spectrum through different scales, from the energy contain wavenumber $\lambda_{\mathcal{F}}^{-1}$ up to the dissipation wavenumber $\lambda_{\mathcal{D}}^{-1}$.}\label{1}
\end{figure}

Figure \ref{1} shows a summary of our theoretical predictions for the magnetic energy spectrum, which emerges from the exact law \eqref{totene4}. This spectrum is roughly consistent with solar wind observations~\citep{S2009}. However, it is worth mentioning that: i) theoretical predictions for the magnetic energy spectrum is strongly dependent of the ratio between magnetic and kinetic energy \citep{M2012,Ba2013}; ii) the actual scaling of the magnetic energy spectra in the solar wind, in particular near and below the electron scale, is still an open question that cannot be resolved unambiguously with the current spacecraft data due to instrumental limitations \citep[e.g.][]{S2013,H2014}. The fate of the turbulent cascade and the resulting dissipation at those small scales is a crucial subject, which is deeply related to the problems {of} particle heating and acceleration in solar wind and in many other astrophysical plasmas \citep{Sc2009}. Our exact results provide a way to estimate the transfer rate of the total energy (and other invariants) over a broad ranges of scales. Application of the exact laws to spacecraft data should therefore inform us about the amount of energy transfered separately into ions and electrons.

Regarding the electron helicity and proton helicity, using definitions \eqref{pote} and \eqref{potp}, equations \eqref{helie3} and \eqref{helip3} can be written as, 
\begin{align}\nonumber
 2\varepsilon_{H_{e}} &=~ \left<\delta(\ue\times\textbf{B})\cdot\delta\textbf{B}\right> -\lambda\mu\left<\delta(\ue\times\ome)\cdot\delta\textbf{B}+\delta(\ue\times\textbf{B})\cdot\delta\ome)\right> \\ \label{helie4} &+ \la^2\mu^2\left<\delta(\ue\times\ome)\cdot\delta\ome\right>, \\ \nonumber
 2\varepsilon_{H_{p}} &=~ \left<\delta(\up\times\textbf{B})\cdot\delta\textbf{B}\right> -\lambda(1-\mu)\left<\delta(\up\times\omp)\cdot\delta\textbf{B}+\delta(\up\times\textbf{B})\cdot\delta\omp)\right> \\ \label{helip4} &+ \la^2(1-\mu)^2\left<\delta(\up\times\omp)\cdot\delta\omp\right>.
\end{align}
When $\mu\rightarrow0$ and $\lambda\rightarrow0$, from both expressions we recover the MHD results for the magnetic helicity \citep{B2016}. Assuming isotropy and a  {maximum} helicity state, the corresponding magnetic helicity spectrum is $E_{H_B}(k)\sim A_kB_k/k\sim k^{-2}$ \citep{P1976}. At HMHD scales, exact scaling laws \eqref{helie4} and \eqref{helip4} correspond to the exact laws for the magnetic and generalized helicity, respectively. In particular, this feature is consistent with the polarization associated with each helicity found recently by \citet{B2016}. Besides, assuming a transfer time $\tau_k^{H_e}\sim(\lambda_ik^2B_k)^{-1}$ for intermediate scales, we obtain $E_{H_e}(k)\sim A_kB_k/k\sim k^{-8/3}$ in the {maximum} helicity state. 

At the smallest {scales where} one can assume $\up\sim0$, equation \eqref{helip4} does not provide new useful information about the proton helicity. On the other hand, the behavior of electron helicity depends strongly on the ratio between magnetic and kinetic {energies}. For instance, at scales proportional to $\lambda_e$ one can assume that the electron kinetic energy is dominant. Since the transfer time is $\tau_k^{H_e}\sim(\mu^2\lambda^2k^3u_{ek})^{-1}$, the electron helicity spectrum corresponds to $E_{H_e}(k)\sim u_{ek}\omega_{ek}/k\sim k^{-8/3}$, which is the same theoretical prediction for the massless EMHD limit \citep{B2016}. In particular, in this scenario where the electron kinetic energy is dominant (and the magnetic energy is negligible), the equation of motion of electrons is similar to the classical 3D hydrodynamic Euler equation (where kinetic energy and kinetic helicity are the two ideal invariants). In this hydrodynamic case, if the large scales of the flow are helical, {there} should be a joint cascade of both energy and helicity to small scales \citep{B1973}. In particular, numerical results strongly support that the magnetic and helicity spectra have the same slope, i.e. $E_U(k)\sim E_H(k)\sim k^{-5/3}$ \citep{C2003a,C2003b}. However, we do not obtain the same slope for both invariants. This result is related to the transfer time of electron helicity at length scales proportional to $\lambda_e=\mu^{1/2}\lambda$ due to the presence of two different species in the plasma. Numerical simulation of 3D incompressible and homogeneous two-fluid plasmas could shed light {onto} some aspects of the nonlinear dynamics of the electron helicity at these smallest scales.

Finally, as we discussed it in the Introduction, in many cases symmetries with preferred directions have direct impact on the structure of exact scaling laws. Therefore, if we consider the presence of a mean magnetic field $\textbf{B}_0$, exact laws \eqref{totene4}, \eqref{helie3} and \eqref{helip3} are modified as,
\begin{align}\nonumber
2\varepsilon_{E_T} &=\frac{1}{\lambda}\big\{\left<\ue'\cdot[\ue\times(\textbf{H}_e+\textbf{B}_0)] + \ue\cdot[\ue'\times(\textbf{H}_e'+\textbf{B}_0)]\right> \\ \nonumber &- \left<\up'\cdot[\up\times(\textbf{H}_p+\textbf{B}_0)] + \up\cdot[\up'\times(\textbf{H}_p'+\textbf{B}_0)\big\}\right> \\  \label{mean1}  &=\frac{1}{\lambda}[\left<\delta(\up\times\textbf{H}_p)\cdot\delta\up\right>-\left<\delta(\ue\times\textbf{H}_e)\cdot\delta\ue\right>], \\ \label{mean2}
2\varepsilon_{H_e} &=~ \left<\delta(\textbf{u}_e\times\textbf{H}_e)\cdot\delta\textbf{H}_e\right> - \big[\left<\textbf{H}_e\cdot(\textbf{u}_e'\times\textbf{B}_0)\right> + \left<\textbf{H}_e'\cdot(\textbf{u}_e\times\textbf{B}_0)\right>\big], \\ \label{mean3}
2\varepsilon_{H_p} &=~ \left<\delta(\textbf{u}_p\times\textbf{H}_p)\cdot\delta\textbf{H}_p\right> - \big[\left<\textbf{H}_p\cdot(\textbf{u}_p'\times\textbf{B}_0)\right> + \left<\textbf{H}_p'\cdot(\textbf{u}_p\times\textbf{B}_0)\right>\big].
\end{align}
As expected, the presence of a local magnetic field does not modify the exact scaling law for the total energy. However, the exact scaling laws associated with the electron helicity and proton helicity are modified by the presence of local magnetic field. This result is compatible with the fact that the presence of a strong magnetic field has a direct impact on the nonlinear dynamics and the turbulent cascade \citep[e.g.][]{GS1995}. Therefore, equations \eqref{mean1}, \eqref{mean2} and \eqref{mean3} may have a wide application for space plasmas, for instance, for the solar wind, which is usually embedded in a moderate uniform magnetic field.

\section{Conclusions}\label{conclus}

We derived exact scaling laws associated with each ideal invariant in a 3D incompressible and homogeneous two-fluid plasma. Without assuming isotropy, we have found exact scaling laws valid in different inertial ranges and independent of the dissipation mechanism present in the plasma. Our main results, i.e. equations \eqref{totene4}, \eqref{helie3} and \eqref{helip3}, are given in term of {two-point increments correlation functions only, which are expressed} in terms on the velocity field of each species and the magnetic field. The data from the recently launched MMS mission have unprecedented high time resolution of the {plasma measurements ($\sim 30$ ms for electrons and $\sim 150$ ms for ions)} should allow us to use the exact laws derived here to analyze the nonlinear cascade in the turbulent plasmas of the magnetosheath and the solar wind, {although the incompressibility assumption may not be valid at sub-ion scales in those media. Furthermore, large statistical samples, i.e. long time series, will be needed at high cadence to ensure the statistical convergence of the estimation of the transfer rate of each particular ideal invariant. This} would give strong {constraints on the theoretical models of turbulence} \citep{I1963,K1965,GS1995}, and could help to study the evolution of spatial anisotropy of the turbulence over a broad range of scales, covering the largest MHD scales to the smallest electron ones.

\section*{Acknowledgments}

NA is supported through an \'Ecole Polytechnique Postdoctoral Fellowship. FS, NA and SG acknowledge financial support from the ANR project THESOW, grant ANR-11-JS56-0008 and from Programme National Soleil-Terre (PNST).

\bibliographystyle{apsrev4-1}

\end{document}